\documentclass[twocolumn,aps,prb,showpacs]{revtex4}
\usepackage{makeidx}
\usepackage{amsmath}
\usepackage{amssymb}
\usepackage{graphicx}
\usepackage{graphics}
\newcommand{\pr}{\parallel}
\newcommand{\pp}{\perp}
\newcommand{\figdir}{}

\newlength{\bilderlength}
\newcommand{\bilderscale}{0.35}
\newcommand{\bilderskip}{\hspace*{0.8ex}}
\newcommand{\diagram}[1]{%
\settowidth{\bilderlength}{\bilderskip%
\includegraphics[scale=\bilderscale]{\figdir#1}\bilderskip}%
\parbox{\bilderlength}{\bilderskip%
\includegraphics[scale=\bilderscale]{\figdir#1}\bilderskip}}
\begin{document}

\title{ \sffamily\bfseries\large Elastic systems with correlated disorder: Response to tilt
and application to surface growth }
\author{ \sffamily\bfseries\normalsize Andrei A. Fedorenko \smallskip}
\affiliation{CNRS-Laboratoire de Physique Th{\'e}orique de l'Ecole Normale Sup{\'e}rieure,%
 24 rue Lhomond, 75231 Paris,  France}

\date{December 5, 2007}


\begin{abstract}
We study elastic systems such as interfaces or lattices pinned by  correlated quenched
disorder considering two different types of correlations: generalized columnar disorder
and quenched defects correlated as $\sim x^{-a} $ for large separation $x$.
Using functional renormalization group methods,
we obtain the critical exponents to two-loop order
and  calculate the response to a transverse field $h$. The correlated disorder
violates the statistical tilt symmetry resulting in nonlinear response to a tilt.
Elastic systems with columnar disorder exhibit a transverse
Meissner effect: disorder generates the critical field $h_c$ below which
there is no response to a tilt and above which the tilt angle behaves
as $\vartheta\sim(h-h_c)^{\phi}$ with a universal exponent $\phi<1$.
This describes the destruction of a weak Bose glass in type-II superconductors
with columnar disorder caused by tilt of the magnetic field.
For isotropic long-range correlated disorder, the linear tilt modulus vanishes
at small fields leading to a power-law response $\vartheta\sim h^{\phi}$
with $\phi>1$. The obtained results are applied  to the Kardar-Parisi-Zhang equation
with temporally correlated noise.
\end{abstract}

\maketitle

\section{ Introduction}
\label{sec1}

\begin{figure}[tbp]
\includegraphics[clip,width=2.0in]{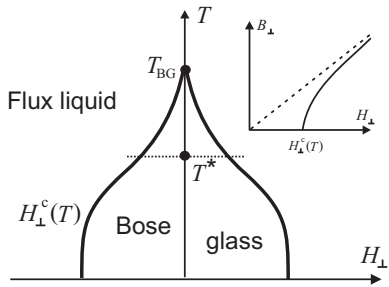}
\caption{Phase diagram of flux lines with columnar disorder at a fixed applied field
$H_{\pr}$ parallel to the columns as a function of transverse field $H_{\pp}$ and
temperature $T$. The Bose glass corresponds to vortex localization and exhibits a
transverse Meissner effect. The transition to liquid phase occurs at the critical field
$H_{\pp}^{c}(T)$. In (1+1) dimensions, the temperatures $T_{\mathrm{BG}}$ and $T^*$
are related to the special values of the Luttinger interaction parameter:
$g(T_{\mathrm{BG}})=3/2$ and $g(T^*)=1$. The inset shows the response to a transverse
field in the vicinity of transition.}
\label{fig-meissner}
\end{figure}

Elastic objects in disordered media are a fruitful concept to study diverse physical systems
such as domain walls in ferromagnets,\cite{domain-walls-exp} charge density waves in solids
(CDW),\cite{cdw} and vortices  in type-II superconductors.\cite{vortex}
In all these systems, the interplay between
elasticity, which tends to keep the object ordered (flat or periodic), and disorder,
which induces distortions, produces a complicated energy
landscape.\cite{fisher-phys-rep98,kardar-phys-rep98,brazovskii03}
This leads to rich glassy behavior.
For instance, at low temperature, weak defects in a crystal of type-II superconductor,
such as oxygen vacancies,  can collectively pin the flux lines  in the so-called Bragg
glass state.\cite{bragg} Vortex pinning prevents the dissipation of energy, and thus, its
understanding has a great importance for applications.
It was observed in experiments that columnar defects produced in the
underlying lattice of superconductors by heavy ion irradiation
can significantly enhance vortex pinning.  \cite{civale91}
Nelson and Vinokur\cite{nelson92} mapped  the problem of flux lines pinned
by columnar defects onto the quantum problem of bosons with uncorrelated
quenched disorder in one dimension less.
The mapping predicts a low temperature ``strong" Bose-glass phase
which corresponds to the localization of bosons in a random potential provided
the longitudinal applied field $H_{\pr}$ is weak enough to create
vortices with density smaller than the density of pins.
For larger $H_{\pr}$, the Bose-glass can coexist with a resistive
liquid of interstitial vortices which, it is  argued, can freeze upon cooling
into a collectively pinned weak Bose-glass phase. \cite{radzihovsky95}
At low tilts of the applied magnetic field relative to the parallel columnar defects,
flux lines remain localized along the defects, so that vortices are
characterized by an infinite tilt modulus.
This phenomenon which is known as the transverse Meissner effect
has been extensively studied experimentally.\cite{smith01}
Vortices undergo a delocalization transition to a flux liquid state at some finite critical
mismatch angle $\vartheta_c$ between the applied field and the direction of defect alignment,
i.e., at some finite transverse field $H_{\pp}^c$.  The schematic phase diagram is shown in
Fig.~\ref{fig-meissner}.
The breakdown of the transverse Meissner effect above $H_{\pp}^c$ can be described by
\begin{equation}
  B_{\pp} \sim (H_{\pp}-H_{\pp}^c)^{\phi},
\end{equation}
where $B_{\pp}$ is the transverse magnetic induction due to the tilted flux lines.
Heuristic arguments of Ref.~\onlinecite{hwa93} based on kink statistics predict
$\phi=1/2$ in $d=1+1$ dimensions and $\phi=3/2$ in $d=2+1$. However,
experiments on a bulk superconductor ($d=3$) with columnar disorder
find $\phi\approx0.5$,\cite{olsson02} while the strong-randomness real-space
renormalization group suggests $\phi=1$ in $d=2$,\cite{refael06} that is in
disagreement  with the predictions based on kink statistics. Thus, further investigations
are needed.

The theoretical advances for elastic objects in  disordered media
are achieved  by developing two general methods: the Gaussian variational approximation
(GVA) and the functional renormalization group (FRG).
GVA relies on the replica method allowing for the replica symmetry breaking.\cite{mezard90}
It is exact in the mean field limit, i.e., in the limit of a large number of components.
FRG is a perturbative renormalization group method
which is able to handle infinite number of relevant operator.\cite{fisher86}
Simple scaling arguments show that the large-scale
properties of a $d$-dimensional  elastic system are governed by uncorrelated disorder
in $d<d_{\mathrm{uc}}=4$. In particular, displacements grow unboundedly with distance,
resulting in a roughness of interfaces or distortions of periodic structures.
The problem is notably difficult
due to  the  so-called dimensional reduction which states that
a $d$-dimensional disordered system at zero temperature is equivalent to all
orders in perturbation theory to a pure system in $d-2$ dimensions at finite temperature.
However, metastability renders the zero-temperature perturbation theory
useless: it breaks down on scales larger than the so-called
Larkin length. \cite{larkin70} The peculiarity of the problem is that
for $d<d_{\mathrm{uc}}$ there is an infinite set of relevant operators.
They can be parametrized by a function which is nothing but the disorder correlator.
The renormalized disorder correlator  becomes a
nonanalytic function beyond the Larkin scale.
\cite{fisher86} The appearance of a
nonanalyticity in the form of a cusp at the origin is related to
metastability, and nicely accounts for the generation of a threshold
force at the depinning transition.\cite{nstl92,lnst07,narayan-fisher93,fedorenko03}
It was recently shown that  FRG
can unambiguously be extended to higher loop order so that the
underlining nonanalytic field theory is probably renormalizable to
all orders. \cite{chauve01,ledoussal02,ledoussal04} Although the two
methods, GVA and FRG, are very different, they provide a fairly consistent picture
and recently a relation between them was established.\cite{ledoussal07}
There is also good agreement with results of
numerical simulations, not only for critical exponents
\cite{roters-pre2002-1,roters-pre2002-2,rosso2003} but also for distributions
of observables\cite{rosso03,fedorenko-fc} and the effective action.\cite{middleton06}

%%%%%%%%%%%%%%%%%%%%%%%%     Figure1. Meissner effect %%%%%%%%%%%%%%%%%%%%
%%%%%%%%%%%%%%%%%%%%%%%%%%%%%%%%%%%%%%%%%%%%%%%%%%%%%%%%%%%%%%%%%%%%%%%%%%%%%%%%

The FRG techniques were also applied to pinning of elastic systems by columnar
disorder.\cite{balents93,chauvembg,chauvethesis} The models studied by FRG, though that
may be more directly applicable to systems such as charge density waves or domain walls,
exhibit many features of the Bose-glass phase of type-II superconductors. In particular,
they demonstrate the absence  of a response to a weak transverse field
and provide a way to compute the exponent $\phi$.\cite{chauvethesis}
However, since FRG intrinsically assumes collective pinning, it also predicts
a slow algebraic decay of translational order, that is not expected in the
strong Bose-glass state when each vortex is pinned by a single columnar pin.
Thus, the FRG is able to handle only a weak Bose-glass phase, exhibiting both
the transverse Meissner effects and the Bragg peaks.

In the present paper, we extend the FRG studies to two-loop order. We also
extend to two-loop order our recent work\cite{fedorenko-pre-2006a} on the elastic
objects in the presence of long-range (LR) correlated disorder with correlations
decaying with distance as a power law. This type of disorder can be induced,
for example, by the presence
of extended defects with random orientations.  In particular, we address the
question of the response to a tilting field  and  compare  the effects
produced by different types of disorder correlations.
The outline of this paper is as follows. Section~\ref{sec2} introduces the models of
elastic objects in the presence of generalized columnar and LR-correlated disorder.
In Sec.~\ref{sec3}, we study the model with LR correlated disorder using FRG up to
two-loop order. In Sec.~\ref{sec4}, we consider the response of elastic objects to a
tilting field and discuss the relation to the quantum  problem of interacting
disordered bosons. In Sec.~\ref{sec5},
we revise the problem of surface growth with temporally correlated noise
using the results obtained in the previous sections.

\section{ Models with correlated disorder}
\label{sec2}

The configuration of elastic object embedded in a $D$-dimensional space
can be parametrized by an $N$-component displacement field $u_{x}$, where $x$
belongs to the $d$-dimensional internal space.
For instance, a $d$-dimensional domain wall corresponds to $d=D-1$ and  $N=1$,
vortices in a bulk superconducter  to $d=D=3$ and $N=2$, and  vortices
confined in a slab to $d=D=2$ and $N=1$.
In this paper, we restrict our study to the case $N=1$ and
elastic objects with  short-range elasticity.
In the presence of disorder, the equilibrium behavior of the elastic object
is defined by the Hamiltonian
\begin{equation}\label{Hamiltonian}
  \mathcal{H}[u] = \int d^d x \left[ \frac{c}2 (\nabla u_x)^2 + V(x,u_x)  \right ],
\end{equation}
where $c$ is the elasticity and $V(x,u)$ is a random Gaussian potential,
with zero mean and variance that will be defined below.
We denote everywhere below $\int_q=\int \frac{d^dq}{(2\pi)^d}$ and
$\int_x=\int d^dx$. The short-scale UV cutoff is implied at $q \sim
\Lambda$ and the system size is  $L$.
The random potential causes the interface to wander
and become rough with displacements growing with the distance $x$ as
$C(x)\sim x^{2\zeta}$. Here, $\zeta$ is the roughness exponent.
Elastic periodic structures lose their strict translational order
and exhibit a slow logarithmic growth of
displacements,  $C(x) = \mathcal{A}_d \ln |x|$.
Although most results of the paper concern the statics at equilibrium,
it is instructive to give a dynamic formulation of the problem.
The driven dynamics of the elastic object in a disordered medium at zero
temperature can be described by the following overdamped equation of motion
\begin{equation} \label{eq-motion}
\eta\partial_t u_{x t} = c\nabla^2 u_{x t} + F(x,u_{x t})+f.
\end{equation}
Here, $\eta$ is the friction coefficient, $F=-\partial_u V(x,u) $ the pinning
force, and $f$ the applied force. The system undergoes
the so-called depinning transition at the critical force $f_c$,
which separates sliding and pinned states.
Upon approaching the depinning transition from the sliding state $f \to f_c^+$
the center-of-mass velocity $v=L^{-d}\int_x\partial_t u_{xt}$
vanishes as a power law
\begin{equation}
  v \sim (f-f_c)^{\beta}.
\end{equation}
In the present work, we consider model (\ref{Hamiltonian}) with two different
types of correlated disorder, which are described in two subsequent sections.

\subsection{Generalized columnar disorder }
\label{sec21}

Real systems often contain extended defects in the form of linear dislocations,
planar grain boundaries, three-dimensional cavities, etc.
We consider the model with extended defects which can be viewed as a
generalization of columnar disorder. The defects are $\varepsilon_d$-dimensional
objects (hyperplanes) extending throughout the whole
system along the coordinate ${x}_{\pr}$ and randomly distributed in the
transverse directions ${x}_{\pp}$  with the concentration taken to be well below
the percolation limit.\cite{dorogovtsev-80,boyanovsky-82,fedorenko-04}
The corresponding correlator of the disorder potential can be written as
\begin{eqnarray}
  \overline{V(x,u)V(x',u')}&=&R(u-u')\delta^{d-\varepsilon_d}(x_{\pp}-x'_{\pp}).
  \label{model-ext}
\end{eqnarray}
The case of  uncorrelated pointlike disorder corresponds to $\varepsilon_d=0$ and
the columnar disorder  to  $\varepsilon_d=1$.
For interfaces, one has to  distinguish two universality classes:
random bond (RB) disorder described by a short-range function $R(u)$ and
random field (RF) disorder corresponding to a function which behaves as $R(u) \sim |u|$
at large $u$. Random periodic (RP) universality class corresponding to a periodic
function $R(u)$ describes systems such as CDW or vortices in $d=1+1$
dimensions.\cite{brazovskii03}

The standard way to average over disorder is the replica trick. Introducing
$n$ replicas of the original system we derive the  replicated
Hamiltonian as follows:
\begin{eqnarray}
 \frac{\mathcal{H}_n[u]}{T}&=&\frac1{2T}\sum\limits_{a}
  \int_x\left[c_{\pr}\left(\nabla_{\pr} u_x^a\right)^2 +c_{\pp}\left(\nabla_{\pp}
   u_x^a\right)^2  \right. \nonumber \\
&& \left. + m^2\left(u_x^a\right)^2 \right]
  -\frac1{2T^2} \sum\limits_{a,b} \int
d^{\varepsilon_d}\,x_{\pr}\,d^{\varepsilon_d}x_{\pr}^{\prime}\,
 \nonumber \\
&& \times d^{d-\varepsilon_d}\,x_{\pp} \
R \left(u_{x_{\pr},x_{\pp}}^a-u_{x_{\pr}^{\prime},x_{\pp}}^b\right), \ \ \ \
\label{H-rep0}
\end{eqnarray}
where we have added a small mass $m$ providing an infrared cutoff.
Replica indices $a$ and $b$ run from $1$ to $n$ and
the properties of the original disordered
system can be restored in the limit $n \to 0$.
We explicitly show in Hamiltonian~(\ref{H-rep0}) that
one has to distinguish the longitudinal and transverse
elasticity modules. Even if the bare elasticity tensor is isotropic, the effective elasticity
may not due to the renormalization by anisotropically distributed disorder.

\subsection{Long-range correlated disorder}
\label{sec22}

In the case of isotropically distributed disorder,
the power-law correlation is the simplest assumption
with possibility for scaling behavior with new fixed points (FPs)
and new critical exponents. The bulk critical behavior of systems
with RB and RF disorder which correlations decay as a power-law $x^{-a}$
was studied in Refs.~\onlinecite{weinrib-83,korucheva-98,fedorenko-00,fedorenko07}.
The power-law correlation of disorder in the $d$-dimensional space
with exponent $a=d-\varepsilon_d$  can be ascribed to
$\varepsilon_d$ dimensional extended defects randomly  distributed
with random orientation. For instance,
$a=d$ corresponds to uncorrelated pointlike
defects, and $a=d-1$ ($a=d-2$) describes infinite lines (planes) of
defects with random orientation.
The power-law correlation with a noninteger value
$a=d-d_f$ can be found in the systems containing fractal-like
structures with the fractal dimension $d_f$. \cite{yamazaki-88}
Here we consider the model with LR-correlated disorder introduced in
Ref.~\onlinecite{fedorenko-pre-2006a} which is defined by the following
disorder correlator:
\begin{eqnarray}
  \overline{V(x,u)V(x',u')}&=&R_1(u-u')\delta^d(x-x')  \nonumber \\
 &+& R_2(u-u')g(x-x'), \label{model-LR}
\end{eqnarray}
with $g(x)\sim x^{-a}$. We fix the constant in the Fourier space
taking $g(q)=q^{a-d}$.
The first term in Eq.~(\ref{model-LR}) corresponds to pointlike disorder with short-range
(SR) correlations  and the second term to LR-correlated disorder.
\textit{A priori} we are interested in the case $a<d$ when the
correlations decay sufficiently slowly, otherwise the disorder is simply SR
correlated.

Using the replica trick, we obtain the replicated Hamiltonian
$\mathcal{H}_n[u]$ and the corresponding action $S[u]$:
\begin{eqnarray}
 \mathcal{S}[u] &=& \frac{\mathcal{H}_n[u]}{T}=\frac1{2T}\sum\limits_{a}
  \int_x\left[c\left(\nabla u_x^a\right)^2 +m^2\left(u_x^a\right)^2\right] \nonumber \\
&&   -\frac1{2T^2} \sum\limits_{a,b} \int_x  R_1(u_x^a-u_x^b)  \nonumber \\
&&   -\frac1{2T^2} \sum\limits_{a,b} \int_{x x'}  R_2(u_x^a-u_{x'}^b)g(x-x').
\label{H-rep1}
\end{eqnarray}
One could start with model (\ref{H-rep1}), setting $R_1(u)=0$.
However, as was shown in Ref.~\onlinecite{fedorenko-pre-2006a}, a nonzero $R_1(u)$
is generated under coarse graining along the FRG flow.
Note that the functions $R_i(u)$ can themselves be SR, LR, or RP.
The generalization of these universality classes to  LR-correlated disorder
is discussed in Ref.~\onlinecite{fedorenko-pre-2006a}.

In the case of uncorrelated disorder, the system (\ref{Hamiltonian}) exhibits
the so-called statistical tilt symmetry (STS), i.e., invariance under transformation
$u_{x} \to u_{x} + f_x$ with an arbitrary function $f_x$. The STS
issues that the one-replica part of the replicated action, i.e., the elasticity,
does not get corrected by disorder to all orders.
The presence of LR-correlated disorder or extended defects destroys the STS,
and thus allows for the renormalization of elasticity.

For a non-Gaussian distribution of disorder, higher order ($p>2$) cumulants would
generate additional terms in the action with  factors of $1/T^p$ and free sums
over $p$ replicas. These terms are irrelevant
in the RG sense that can be seen by power counting, and thus will be neglected
from the beginning.

\section{ Renormalization of the model with long-range correlated disorder}
\label{sec3}

\subsection{Perturbation theory and diagrammatics}
\label{sec31}

We now study the scaling behavior of model (\ref{H-rep1})
starting with simple power counting. The elastic term in action (\ref{H-rep1})
is invariant under $x \to x b $,  $u \to u  b^{\zeta} $, $c \to b^{-\psi} c$
provided  $T \to b^{\theta_T} T$ with $\theta_T=d-2+2\zeta-\psi$.
Since $\theta_T$ is positive near $d=4$ the temperature $T$ is formally
irrelevant.
The STS would fix $\psi=0$, however, this is not the case here.
$\zeta$ and $\psi$ are for now undetermined and their actual values
will be fixed by the disorder correlators at the stable FP.
Under the rescaling transformation the disorder correlation functions $R_1$ and $R_2$
go up by factors $b^{d-2\theta_T}=b^{4-d-4\zeta+2\psi}$ and
$b^{d-2\theta_T}=b^{4-a-4\zeta+2\psi}$, respectively.
Thus, in the vicinity of Gaussian FP ($R_i=0$),
SR disorder becomes relevant for $\zeta-\psi/2<(4-d)/4$  and LR disorder
is naively relevant for  $\zeta-\psi/2<(4-a)/4$. \textit{A posteriori} these inequalities
are satisfied at the RB and RP FPs. For RF disorder, however,
power counting suggests that  SR disorder is relevant for
$\zeta-\psi<(4-d)/2$, while LR disorder
is relevant for $\zeta-\psi<(4-a)/2$.\cite{fedorenko-pre-2006a}

%%%%%%%%%%%%%%%%%%%%%%%%%%%%%%%%%%%%Figure2. Disorder %%%%%%%%%%%%%%%%%%%%%%%%%%%%%%%%%%%%%%%%
\begin{figure}[tbp]
\includegraphics[clip,width=2.7in]{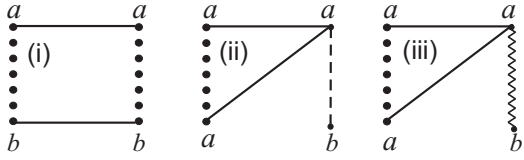}
\caption{Two-replica one-loop diagrams correcting disorder. The dot line corresponds to
either SR disorder vertex (dashed line) or to LR disorder vertex (wavy line). Diagrams of types
 (i) and (ii) contribute to SR disorder.  Only diagrams of type (iii) give corrections
 to LR disorder. }
\label{fig-disorder}
\end{figure}
%%%%%%%%%%%%%%%%%%%%%%%%%%%%%%%%%%%%%%%%%%%%%%%%%%%%%%%%%%%%%%%%%%%%%%%%%%%%%%%%

Let us consider the perturbation theory in disorder and its diagrammatic representation.
In momentum space, the quadratic part of action (\ref{H-rep1}) gives rise to
the free propagator
$\langle u^a_q u^b_{q'} \rangle_0 = (2\pi)^d \delta^d(q+q')T \delta_{ab}C(q)$
represented graphically  by a line:
\begin{equation}
{}^{a}\frac{\hspace{15mm}}{\hspace{15mm}}^{b}
 \ \ \ \ =  T C(q)\delta_{ab}=\frac{T}{c q^2 +m^2}\delta_{ab}. \label{C}
\end{equation}
We will distinguish two different interactions, SR and LR,
for which we adopt the following splitted diagrammatic representation:
\begin{eqnarray}
  \diagram{vertexSR} &=&
 \sum\limits_{ab} \frac{R_1(u_x^a-u_{x}^b)}{2T^2}, \\
\diagram{vertexLR} &=&
 \sum\limits_{ab} \frac{R_2(u_x^a-u_{x'}^b)}{2T^2}g(x-x').
\end{eqnarray}
Following the standard field theory renormalization program,
we compute the effective action and determine counter-terms to render the theory
UV finite as $d,a\to 4$. To regularize integrals, we use
a generalized dimensional regularization
with a double expansion in $\varepsilon=4-d$ and $\delta=4-a$.
The effective action $\Gamma[u]$ is defined by
the Legendre transform $\Gamma[u]=J u - W[J]$, $\mathcal{W}'[J]=u$ of the
generating functional for connected correlators $\mathcal{W}[J]=\ln \mathcal{Z}[J]$.
The replicated partition function $\mathcal{Z}$ in the presence of sources $J$ is given by
\begin{equation}
  \mathcal{Z}[J]=\int \prod\limits_a \mathcal{D}u_a\, \exp\left(
  -\mathcal{S}[u]+\int_x\sum\limits_a J_x^a u_x^a   \right).
\end{equation}
The effective action $\Gamma[u]$ is by definition a generating functional of
one-particle irreducible vertex functions. However, it turns out to be nonanalytic in
some directions, and therefore, the relying on the expansion in $u$ is danger.
To overcome these difficulties, we
employ the formalism of functional diagrams introduced in Ref.~\onlinecite{ledoussal04}.
Since the temperature is formally irrelevant, we compute the correction to
the effective action at $T=0$.
Analyzing UV divergences of the functional diagrams contributing
to the effective action, we find
that the disorder is corrected only by local parts of two-replica diagrams
and the elasticity only by one-replica diagrams.

\subsection{ Correction to disorder and $\beta$ functions}
\label{sec32}

To one-loop order at $T=0$, the correction to disorder is given by the
local parts of the two-replica diagrams shown in Fig.~\ref{fig-disorder}.
The corresponding expressions read
\begin{eqnarray}
  \delta^1 R_1(u) &=& \left[ \frac12 R_{10}''(u)^2-R_{10}''(u)R_{10}''(0) \right]I_1 \nonumber \\
  &&  +  \left[ R_{10}''(u) R_{20}''(u)-R_{10}''(u)R_{20}''(0)\right]I_2 \nonumber \\
  &&  +   \frac12 R_{20}''(u)^2 I_3, \ \ \ \ \ \ \\
\delta^1 R_2(u) &=& - R_{20}''(u)R_{10}''(0)I_1 - R_{20}''(u)R_{20}''(0)I_2,\ \ \ \ \
\end{eqnarray}
where we have included factor of $1/c_0^2$ in $R_{i0}(u)$. In this section,
bare parameters are denoted by the subscript ``0".
The one-loop integrals $I_1$, $I_2$ and $I_2$  diverge logarithmically
and for $\varepsilon,\delta \to 0$ are given by
\begin{eqnarray}
  I_1&=&\int_q \frac1{(q^2+\hat{m}^2)^2} = K_4 \frac{\hat{m}^{-\varepsilon}}{\varepsilon}
   + \mathcal{O}(1), \label{I1} \\
  I_2&=&\int_q  \frac{q^{a-d}}{(q^2+\hat{m}^2)^2}=K_4 \frac{\hat{m}^{-\delta}}{\delta}
   + \mathcal{O}(1),\label{I2} \\
  I_3&=&\int_q  \frac{q^{2(a-d)}}{(q^2+\hat{m}^2)^2} =\frac{K_4
  \hat{m}^{-2\delta+\varepsilon}}{2\delta-\varepsilon} + \mathcal{O}(1),
  \end{eqnarray}
where we have set $\hat{m}=m/\sqrt{c_0}$ and $K_d$ is the area of
a $d$-dimensional sphere divided by $(2\pi)^d$.
Let us define the renormalized dimensionless disorder $R_{i}$ as
\begin{eqnarray}
 {m}^{\varepsilon} R_{1}(u)=R_{10}(u)+\delta^1 R_1(u), \\
 {m}^{\delta} R_{2}(u)=R_{20}(u)+\delta^1 R_2(u).
\end{eqnarray}
Note that to one-loop order, there is no correction due to the renormalization of elasticity
(see below).
The $\beta$ functions are defined  as the derivative of $R_i(u)$
with respect to the mass $m$ at fixed bare disorder $R_{i0}(u)$. It is convenient to
rescale the field $u$ by $m^{\zeta}$ and write the $\beta$ functions for
the function $\tilde{R}_i=K_4 m^{-4\zeta}R_i(um^{\zeta})$.
Dropping the tilde subscript, the flow equations to one-loop order read
\begin{eqnarray}
\partial_{\ell} R_1(u) &=&(\varepsilon-4\zeta) R_1(u) + \zeta u R'_1(u) \nonumber \\
&&+ \frac12 [R_1''(u)+R_2''(u)]^2 + A R_1''(u), \label{frg-R-1} \\
\partial_{\ell} R_2(u) &=&(\delta-4\zeta)R_2(u) + \zeta u R'_2(u) +   A R_2''(u),\qquad
\label{frg-R-2}
\end{eqnarray}
where $A=-[R_1''(0)+R_2''(0)]$ and
$\partial_{\ell} := - m \frac{\partial}{\partial {m}}$.
The FPs of flow equations (\ref{frg-R-1}) and (\ref{frg-R-2})
characterizing different universality classes have been computed numerically
in Ref.~\onlinecite{fedorenko-pre-2006a} and the
corresponding critical exponents have been derived to first order
in $\varepsilon$ and $\delta$. The remarkable property of the FRG flow  is that the
LR part of disorder correlator $R_2(u)$ remains an analytic function along the flow
for all universality classes. We will show below that due to this feature, one can
obtain the critical exponents to two-loop order just computing the two-loop correction to
elasticity and avoiding exhaustive two-loop calculations.

\subsection{ Correction to elasticity}
\label{sec33}

The STS violation causes a renormalization of elasticity.
The first order correction to the single-replica part of
effective action is expressed by the following diagram:
\begin{equation}
\!\! \diagram{elasticity} = - \int_{x,x'}
 \sum\limits_a \frac{R_{20}''(u_x^a-u_{x'}^a) g(x-x')}{2T} C (x-x'),
\end{equation}
where the bare correlation function $C(x)$ is given by Eq.~(\ref{C}).
Using the short distance expansion
\begin{equation} \label{short}
  u_x^a-u_{x'}^a = \sum\limits_{i=1}^d (x_i-x'_i)\frac{\partial u^a_x}{\partial x_i}+...
\end{equation}
and identifying the terms of the kind $-(\nabla u_x^a)^2/2T$ as a correction
to elasticity, we find
\begin{eqnarray}
  \delta^1 c &=& \frac1{2d} R_{20}^{(4)}(0)\int_x x^2 g(x)C(x) \nonumber \\
& =& c_0  R_{20}^{(4)}(0) \frac{\delta-\varepsilon}{4\delta} \hat{m}^{-\delta}
+ O(\varepsilon,\delta),
 \label{del1}
\end{eqnarray}
where in the last line we have included $K_4/c_0^2$ in a redefinition of  $R_{20}(u)$.
Since Eq.~(\ref{del1}) is finite for
$\varepsilon, \delta \to 0$, the elasticity does not get corrected to one-loop order.

%%%%%%%%%%%%%%%%%%%%%%%%%%%%%% Figure 3. Elasticity 2-loop %%%%%%%%%%%%%%%%%%%%%%
\begin{figure}[tbp]
\includegraphics[clip,width=2.7in]{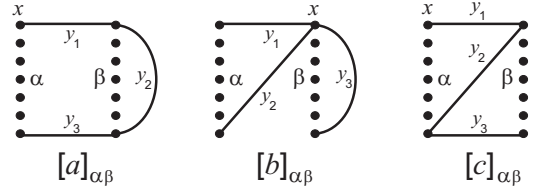}
\caption{Single-replica two-loop diagrams correcting elasticity.
The dotted  line corresponds to either SR disorder vertex ($\alpha,\beta=1$)
or to LR disorder vertex ($\alpha,\beta=2 $). The solid line corresponds to  bare correlation
functions $C(y)$. The corresponding expressions are computed in the Appendix. }
\label{fig-2loop}
\end{figure}
%%%%%%%%%%%%%%%%%%%%%%%%%%%%%%%%%%%%%%%%%%%%%%%%%%%%%%%%%%%%%%%%%%%%%%%%%%%%%%%%

We now turn to the two-loop corrections. The three different sets of diagrams
contributing to elasticity are depicted in Fig.~\ref{fig-2loop}.
The details of calculations are given in the Appendix.
Summing up all contributions, we arrive at
\begin{eqnarray}
  \frac{\delta^2 c}{c_0} &=&
R^{\prime\prime}_{10}(0)R^{(6)}_{20}(0) \hat{m}^{-(\varepsilon+\delta)}
\frac{\varepsilon-\delta}{\varepsilon+\delta}
\frac1{4\varepsilon} \nonumber \\
&&+ R^{\prime\prime}_{20}(0)R^{(6)}_{20}(0)  \hat{m}^{-2\delta}
\frac{\varepsilon-\delta}{8 \delta^2} + O(1). \label{del2}
\end{eqnarray}
To render the poles in $\varepsilon$ and $\delta$, we introduce
the renormalization group Z factor as follows:
\begin{equation}
   c=Z_c[R_1,R_2]^{-1} c_0. \label{Z}
\end{equation}
The exponent $\psi$ is given then by
\begin{eqnarray}
  \psi &=&\left. -{m}\frac{d}{d {m}} \ln Z_{c} [R_1,R_2]\right|_0, \label{c}
\end{eqnarray}
where subscript 0 indicates a derivative at constant bare parameters.
Taking the derivative with respect to the mass, we obtain
\begin{eqnarray}
 \left.  -{m}\frac{d}{d {m}} \ln Z_{c} \right|_0 &=&
 - \frac14 R_{20}^{(4)}(0) ({\delta-\varepsilon}) \hat{m}^{-\delta} \nonumber \\
 && +
R^{\prime\prime}_{10}(0)R^{(6)}_{20}(0) \hat{m}^{-(\varepsilon+\delta)}
\frac{\delta-\varepsilon}{4\varepsilon} \nonumber \\
&&+ R^{\prime\prime}_{20}(0)R^{(6)}_{20}(0)  \hat{m}^{-2\delta}
\frac{\delta-\varepsilon}{4 \delta}. \label{psi-bare}
\end{eqnarray}
To calculate $\psi$, we have to express the bare disorder via renormalized one
as follows:
\begin{eqnarray}
  R_{20}^{(4)}(0)&=&m^{\delta}\left[R_2^{(4)}(0)
  +R_1^{\prime\prime}(0)R_2^{(6)}(0)\frac1{\varepsilon} \right. \nonumber \\
  &&+ \left. R_2^{\prime\prime}(0)R_2^{(6)}(0)\frac1{\delta} \right]. \label{bare}
\end{eqnarray}
Substituting Eq.~(\ref{bare}) in Eq.~(\ref{psi-bare}), we find that the leading two-loop
corrections are exactly canceled  by the counter-terms, so that we leave with
\begin{eqnarray}
  \psi  &=& - \frac14 (\delta-\varepsilon ) R^{(4)}_2(0). \label{psi-2}
\end{eqnarray}
The finite part of the single-replica two-loop diagrams (\ref{del2}) is expected to correct
elasticity at three-loop order. Hence, we argue that the perturbation theory for this model
is organized in such a way that the single-replica $p$-loop diagrams correct the elasticity
only to $(p+1)$ order. Since $R_2(u)$ remains analytic along the FRG flow, we have
$R^{(4)}_2(0)>0$, and therefore, $\psi<0$. The corresponding values
of the exponent $\psi$ computed for the RF, RB, and RP universality classes using
the FPs found in Ref.~\onlinecite{fedorenko-pre-2006a} are shown in Fig.~\ref{fig-psi}.

\subsection{Roughness exponent to two-loop order}
\label{sec34}

%%%%%%%%%%%%%%%%%%%%%%%%     Figure4. Exponent Psi %%%%%%%%%%%%%%%%%%%%
\begin{figure}[tbp]
\includegraphics[clip,width=3.2 in]{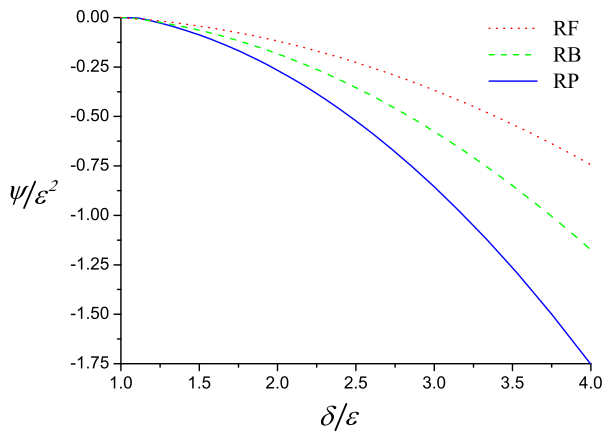}
\caption{(Color online) Exponent $\psi$ as a function of  $\delta/\varepsilon$ at two-loop order
for the RF, RB, and RP universality classes.}
\label{fig-psi}
\end{figure}
%%%%%%%%%%%%%%%%%%%%%%%%%%%%%%%%%%%%%%%%%%%%%%%%%%%%%%%%%%%%%%%%%%%%%%%%%%%%%%%%

We now  show how one can calculate the roughness exponent $\zeta$ to second order
in $\varepsilon$ and $\delta$ knowing only the exponent $\psi$
computed to second order in Sec.~\ref{sec33}. To that end, we do not need the whole
FRG to two-loop order. Let us start with the RB universality class.
The roughness exponent is fixed by a stable RB FP solution
of Eqs.~(\ref{frg-R-1}) and (\ref{frg-R-2}) which decays exponentially fast for large $u$.
The equations possess both the SR RB FP with $R_2(u)=0$
and the LR RB FP with $R_2(u)\ne 0$.
The roughness exponent corresponding to the SR RB FP is known to second order
in $\varepsilon$ and reads\cite{chauve01,ledoussal04}
\begin{equation}
  \zeta_{\mathrm{SRRB}}=0.208298\varepsilon+
   0.006858\varepsilon^2 + {O}(\varepsilon^3). \label{psi-rb}
\end{equation}
Despite the smallness of the two-loop correction, the estimation
of the exponent in $d=1$,
$\zeta_{\mathrm{SRRB}}=0.6866$
given by  Eq.~(\ref{psi-rb}), visibly differs from the known exact result $2/3$.
One can improve the accuracy of $\zeta$ by use the
Pad\'{e} approximant [2/1] involving also the unknown third order correction.
Tuning the latter in order  to reproduce the exact result $2/3$ for
$\varepsilon=3$, we end up with the expression
\begin{equation}
  \zeta_{\mathrm{SRRB}}^{\mathrm{imp}}=
  \frac{0.208298\varepsilon + 0.040017\varepsilon^2}{1+0.159192\varepsilon}, \label{psi-rb-im}
\end{equation}
which is expected to be fairly accurate for $0\le\varepsilon\le3$.

%%%%%%%%%%%%%%%%%%%%%   Figure5. Stability of RB fixed points %%%%%%%%%%%%%%%%
\begin{figure}[tbp]
\includegraphics[clip,width=3.2 in]{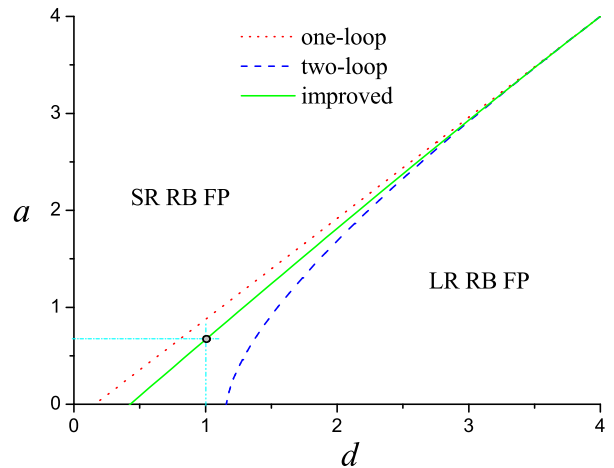}
\caption{(Color online) Stability regions of SR and LR RB FPs on plane $(d,a)$. The
borderline between regions is given by equation $\zeta_{\mathrm{LRRB}}=\zeta_{\mathrm{SRRB}}$
(on the exact crossover line also $\psi=0$).
The shown lines are computed using the one-loop, two-loop, and improved results. The circle
is a point on the exact crossover line.}
\label{fig-stab}
\end{figure}
%%%%%%%%%%%%%%%%%%%%%%%%%%%%%%%%%%%%%%%%%%%%%%%%%%%%%%%%%%%%%%%%%%%%%%%%%%%%%%%%

\begin{figure}[tbp]
\includegraphics[clip,width=3.2 in]{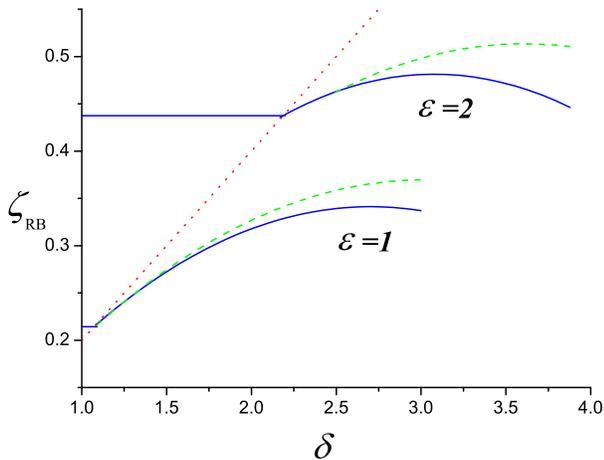}
\caption{(Color online) RB disorder: roughness exponent as a function of $\delta$ for
$\varepsilon=1$ and $\varepsilon=2$. The solid lines are computed using the improved
estimation  (\ref{psi_im}) and (\ref{f2}) of the exponent $\psi$.
The dashed lines are computed using $\psi$ which is given by Eq.~(\ref{psi-2}) and
shown in Fig.~\ref{fig-psi}. The dotted line is the
one-loop result $\zeta_{\mathrm{LRRB}}=\delta/5$
which does not depend on $\varepsilon$.}
\label{fig-rough}
\end{figure}

We now focus on the LR RB FP with $R_2(u)\neq 0$. We can integrate
both sides of flow equation~(\ref{frg-R-2}) over $u$ from $0$ to $\infty$
taking into account that for RB  disorder $R_2(u)$ decays exponentially fast.
Since for RB disorder the integral $\int_0^{\infty} du\,R_2(u)$ is nonzero, we can
determine  the roughness exponent $\zeta_{\mathrm{LRRB}}=\delta/5$ to first order
in $\varepsilon$ and $\delta$.
Fortunately, one can go beyond the one-loop
approximation. Indeed, the direct inspection of diagrams contributing to
the flow equation~(\ref{frg-R-2}) shows that the higher orders can only be
linear in even derivatives of $R_2(u)$.  The only term which is linear in $R_2(u)$
comes from the renormalization of elasticity and can be rewritten
as $2\psi R_2(u)$ to all orders. Hence, we have to all orders
\begin{eqnarray}
\partial_{\ell} \int_0^{\infty} du R_2 (u)
&=& (\delta - 5 \zeta +2\psi)\int_0^{\infty} du R_2(u) \label{flow-int-r2-2}, \ \ \
\end{eqnarray}
and as a consequence, $\int_0^{\infty} du R_2(u)$ is
exactly preserved along the FRG flow resulting in the exact identity
\begin{equation}\label{zeta LRRB-2}
\zeta_{\mathrm{LRRB}}=\frac{\delta + 2\psi}5.
\end{equation}
Substituting Eq.~(\ref{psi-2}) into Eq.~(\ref{zeta LRRB-2}),
we obtain the roughness exponent $\zeta_{\mathrm{LRRB}}$
to second order in $\varepsilon$ and $\delta$.
Before we proceed to compute the exponents, let us to check stability of the SR and LR RB FPs.
As was shown in Ref.~\onlinecite{fedorenko-pre-2006a}, the SR RB FP is
unstable with respect to LR disorder if
$\zeta_{\mathrm{LRRB}}>\zeta_{\mathrm{SRRB}}$. To one-loop order, this gives that
the SR RB FP is stable for $\delta<1.0415\varepsilon$.
Equating (\ref{zeta LRRB-2}) and (\ref{psi-rb-im}), we can compute the stability
regions to second order in $\varepsilon$ and $\delta$ (see Fig.~\ref{fig-stab}).
The alternative way to determine the crossover line relies on the requirement that
the exponent $\psi$ is a continuous function of $\varepsilon$ and $\delta$.
It is zero in the region controlled by the SR FP, and therefore has to vanish
when approaching the crossover line from the LR stability region.
Since the $\psi$ is of second order in $\varepsilon$ and $\delta$,
the $\psi$ criterion at two-loop order gives the same stability regions
as the roughness exponents equating at one-loop order.
However, we can significantly improve the latter if we take into account that
$\psi=0$ on the crossover line. The resulting crossover line is shown in Fig.~\ref{fig-stab}.
We can also improve the two-loop estimation of $\psi$.
To that end, we write down a formal expansion of $\psi$ in $\varepsilon$,
\begin{equation} \label{psi_im}
  \psi = \varepsilon^2 f_1(\delta/\varepsilon)+
          \varepsilon^3 f_2(\delta/\varepsilon)+...
\end{equation}
The function $f_1(x)$ is basically the function shown in Fig.~\ref{fig-psi}.
We now tune the function $f_2(x)$
in order to make $\psi=0$ on the crossover line and find
\begin{eqnarray} \label{f2}
  f_2(x)=\frac{0.159192x-0.200087}{x-1.04149}f_1(x).
\end{eqnarray}
Using Eqs.~(\ref{psi_im}) and (\ref{f2}),  we compute the roughness exponent
$\zeta_{\mathrm{LRRB}}$ as a function of $\delta$ for $\varepsilon=1$
and $\varepsilon=2$ (see Fig.~\ref{fig-rough}).  Unfortunately, the accuracy
rapidly decays with $\varepsilon$, so that estimation of the roughness exponent for
$\varepsilon=3$ is very difficult and postponed to  Sec.~\ref{sec5}.

Similar to the case of RB disorder, one can show that the roughness exponent at the LR RF
FP is exactly given by
\begin{equation}
\zeta_{\mathrm{LRRF}}=\frac{\delta + 2\psi}3,
\end{equation}
and the crossover line between the SR and LR RF FPs is exactly given by $\delta=\varepsilon$.

%%%%%%%%%%%%%%%%%%%%%   Figure6. Roughness for the RB fixed point %%%%%%%%%%%%%%%%
%%%%%%%%%%%%%%%%%%%%%%%%%%%%%%%%%%%%%%%%%%%%%%%%%%%%%%%%%%%%%%%%%%%%%%%%%%%%%%%%

\section{Response to tilt}
\label{sec4}

In this section, we study the response of a $d$-dimensional elastic object to a small
tilting force tending to rotate the object in the plane $(x_1,u)$. The tilting force
can be incorporated into the Hamiltonian as follows:
\begin{equation}
   \mathcal{H}_{h}[u]= \mathcal{H}[u]-h \int d^d x\,  \partial_{x_1} u_x.
\end{equation}
Such a force can be caused, for example, by a tilt of the applied field
in superconductors or by tilted boundary conditions in the case of interfaces.
For superconductors, we have $h=\phi_0 H_{\pp}/(4\pi)$, where
$H_{\pp}$ is the component of the applied magnetic field transverse to the
flux lines directed along $x_1$ and $\phi_0$ is the magnetic flux quantum.\cite{hwa93}
Since we restrict our consideration to the case $N=1$, our results can be applied
only to flux lines confined in $(1+1)$ dimensions. However, the methods we use here
can be extended to general $N$, and therefore applied to vortices in $(2+1)$ dimensions.

We focus on the response of the system to a small field $h$,
which can be measured by the average angle between the perturbed and unperturbed orientations
of the object in the $(x_1,u)$ plane: $\vartheta(h):= \overline{\partial_{x_1} u_x }$.
In the absence of disorder the straightforward minimization of the Hamiltonian leads
to the linear response: $\vartheta(h) = h/c$.
To study the effect of disorder, it proves more convenient to work in the tilted
frame:  $u_x \to u_x + \vartheta x_1$. The corresponding Hamiltonian is
\begin{eqnarray}
   \mathcal{H}_{h}[u] &=& \int d^d x\,\left[ \frac12 \sum\limits_{i=1}^d
   {c_i} (\partial_{x_i} u_x)^2 + V(x,u_x+\vartheta x_1) \right. \nonumber \\
   &&  -(h-c_1 \vartheta)  \partial_{x_1} u_x \Big], \label{H-tilt}
\end{eqnarray}
where the field $u$ satisfies $\overline{\partial_{x_1} u_x }=0$.
Note that due to the violation of the STS symmetry, the tilted system can exhibit
anisotropic effective elasticity even if the bare elasticity and disorder are isotropic.
We now show by simple power counting that a finite tilt does introduce a new length
scale in the problem which can be associated with the correlation length defined
through the connected two point correlator,
\begin{equation}
  \Omega(x; \vartheta)=\overline{\partial_{x_1} u(0)\partial_{x_1} u(x)}^c \sim
 \exp({- x_1/\xi_{\vartheta}}).
\end{equation}
Indeed, upon scaling transformation $x\to b x $, $u \to b^{\zeta} u$
the arguments of the disorder term in Hamiltonian (\ref{H-tilt}) scale like
$V(bx, b^{\zeta} u_x  + \vartheta b x_1 )$. Comparing two terms of the last argument,
we find that finite $\vartheta$ changes
the character of disorder correlator above the length scale
\begin{equation} \label{xi-def}
  \xi_{\vartheta}\sim \vartheta^{-1/(1-\zeta)},
\end{equation}
diverging for $\vartheta \to 0$ provided that $\zeta<1$.
Below $\xi_{\vartheta}$
one can neglect the tilt, while above $\xi_{\vartheta}$ the dependence on $u_x$
is completely washed out and the $\vartheta$ term starts to suppress the correlation
of disorder along $x_1$. Thus, $\xi_{\vartheta}$ serves as the correlation length
along $x_1$, and therefore $c_1$ does not get
renormalized beyond this scale.
In the next two sections, we investigate the difference in the response to tilt for
anisotropically distributed extended defects and isotropic LR-correlated disorder.

\subsection{Response in the presence of columnar disorder}
\label{sec41}

Here, we extend the previous one-loop FRG
studies\cite{balents93,chauvethesis} of elastic systems in the presence of
columnar disorder to two-loop order and proceed to describe the
transverse Meissner physics in a quantitative way.
We consider the model with $\varepsilon_d$-dimensional extended defects
introduced in Sec.~\ref{sec21}.
We take $c_i=c_{\pr}$ $(i=1,...,\varepsilon_d)$
and we are also free to put $c_i=1$ $(i=\varepsilon_d+1,...,d)$ since they do not
get corrected by disorder. Simple power counting shows that the upper critical dimension
of the problem is $d_{\mathrm{uc}}=4+\varepsilon_d$.
We use the dimensional regularization of integrals with
a $\tilde{\varepsilon}=4-d+\varepsilon_d$ expansion.
The FRG flow equations to two-loop order read
\begin{eqnarray}
&& \partial_{\ell} R(u) = (\tilde{\varepsilon}-4\zeta)R(u)+ \zeta u R'(u) - TR^{(4)}(u)
    \nonumber \\
&& \hspace{12mm}+ \frac12 R''(u)^2 -R''(0) R''(u)-\frac12 R'''(0^+)^2R''(u) \nonumber \\
&& \hspace{12mm}+ \frac12 (R''(u)-R''(0)) R'''(u)^2, \label{ren-dis} \\
&&\partial_{\ell} \ln c_{\pr} =  R^{(4)} (0) +
   R^{(4)}(0)^2 + 2 R'''(0)R^{(5)}(0), \label{ren-c}\ \ \ \\
&& \partial_{\ell} \ln T = -\theta_T-\frac{\varepsilon_d}2R^{(4)}(0) + O(R^2),
   \label{ren-T}\ \ \ \\
&& \partial_{\ell} \tilde{h} = c_{\pr}^{1/2} \Lambda_0 e^{-\ell} R'''(0^ +) + O(R^2).
 \label{ren-h}
\end{eqnarray}
where $\theta_T=d-2+2\zeta$ and  $\Lambda_0$ is the bare cutoff. In Eq.~(\ref{ren-h}),
$\tilde{h}$ is the coefficient in front of the term
\begin{equation}\label{h-tilde}
 \tilde{h} \sum\limits_a \int d^dx\, |\nabla_{\pr}u_x^{a}|
\end{equation}
ultimately generated in the effective Hamiltonian along the FRG flow.\cite{balents93}
The correction to $\tilde{h}$ is strongly UV diverging, and thus is nonuniversal.
Note that the flow equation for the  generalized columnar disorder (\ref{ren-dis}) coincides
to all orders with that for pointlike disorder up to change
$\tilde{\varepsilon}\to \varepsilon$.

%%%%%%%%%%%%%%%%%%%%%   Figure6. Response     %%%%%%%%%%%%%%%%
\begin{figure}[tbp]
\includegraphics[clip,width=3.2 in]{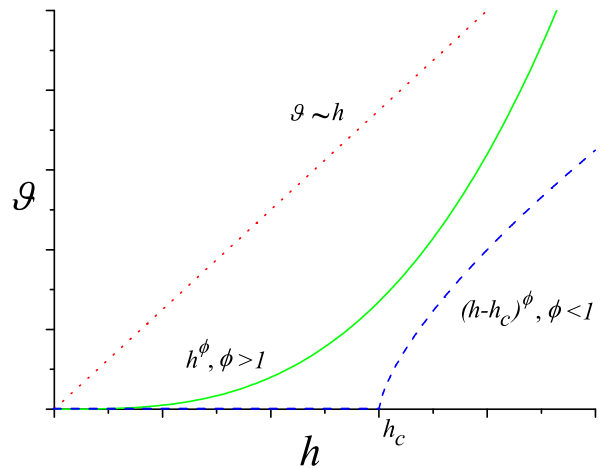}
\caption{ (Color online) Schematic  plot of the elastic object
response to a transverse field for uncorrelated disorder (dotted line),
columnar disorder (dashed line), and LR-correlated disorder (solid line).}
\label{fig-resp}
\end{figure}
%%%%%%%%%%%%%%%%%%%%%%%%%%%%%%%%%%%%%%%%%%%%%%%%%%%%%%%%%%%%%%%%%%%%%%%%%%%%%%%%

Let us start from the analysis  at $T=0$.
The flow picture very resembles that for the depinning transition with  tilt $\vartheta$,
longitudinal elasticity $c_{\pr}$, and tilting field $h$ playing the roles
of  velocity, friction, and driving force, respectively.
For $d<4+\varepsilon_d$, the running disorder correlator $R^{(4)}(u)$
blows up at the Larkin scale
\begin{equation}
  l_c=\frac1{\tilde{\varepsilon}} \ln\left( 1+
  \frac {\tilde{\varepsilon}}  {3|R_0^{(4)}(0)|}  \right).
\end{equation}
Consequently, the longitudinal elasticity diverges at zero tilt $\vartheta$
in a way  similar to mobility divergence at the depinning transition in quasistatic limit.
Beyond the Larkin scale $l>l_c$, the $R^{(2)}(u)$ develops a cusp at origin, $R'''(0^+)>0$.
The term (\ref{h-tilde}) is generated in the effective Hamiltonian and the
$R^{(4)}(0)$ changes its sign from positive to negative. The latter
leads to a power law decay of the longitudinal elasticity $c_{\pr}\propto L^{-\psi}$
with
\begin{equation}
  \psi = -R^{*(4)}(0)-R^{*(4)}(0)^2 - 2 R^{*\prime\prime\prime}(0)R^{*(5)}(0),
\end{equation}
where $R^*(u)$ is a FP solution of the flow equation (\ref{ren-dis}).
Similar to the threshold force generation at the depinning transition,
term (\ref{h-tilde}) reduces the tilting force and
generates the critical tilting force $h_c$.
The flow equation (\ref{ren-h}) allows us to estimate the nonuniversal value of $h_c$.
Integrating Eq.~(\ref{ren-h}) up to large scales, we find
\begin{equation}
  h_c = \frac{c_{0}^{1/2}\Lambda_0e^{-l_c(1+\psi/2)}}{1+\psi/2} R^{*\prime\prime\prime}(0+).
\end{equation}
We now in a position to compute the exponent $\phi$, which we define as
\begin{equation}
  \vartheta\propto(h-h_c)^{\phi}.
\end{equation}
To that end, we renormalize the equilibrium balance equation $h-h_c = c_1(L) \vartheta$
up to the scale $L=\xi_{\vartheta}$ at which the elasticity $c_1$ stops to get renormalized.
Using Eq.~(\ref{xi-def}), we obtain the exact scaling relation
\begin{equation}
  \phi= \left(1+\frac{\psi}{1-\zeta}\right)^{-1}. \label{phi}
\end{equation}
The exponents $\zeta$, $\psi$, and  $\phi$ computed to second order in $\tilde{\varepsilon}$
for different universality classes are summarized in Tab.~\ref{table1}.
Note that expansions in $\tilde{\varepsilon}$ are expected to be
Borel nonsummable, and  thus ill behaved for high orders  and
large $\tilde{\varepsilon}$. In this light, the
using of exact relation (\ref{phi}) may be more favorable than
the expansions given in the last column of Table~\ref{table1}.
Systems described by the RP universality class
exhibit  slow logarithmic growth  of displacements
\begin{eqnarray}
\overline{(\bar{u}_{x_{\pp}}-\bar{u}_0)^2}= \mathcal{A}_d \ln x_{\pp},
\end{eqnarray}
where $\bar{u}_{x_{\pp}}:=L^{-\varepsilon_d} \int d^{\varepsilon_d}x_{\pr}\, u_{x}$.
The universal amplitude can be easily deduced from the results for the uncorrelated
disorder and to two loop reads
\begin{eqnarray}
\mathcal{A}_d = \frac{\tilde{\varepsilon}}{18} + \frac{\tilde{\varepsilon}^2}{108}
+ O(\tilde{\varepsilon}^3),
\end{eqnarray}
where we have fixed the period to $1$. The logarithmic growth of displacements corresponds
to a slow power-law decay of the translation order, and thus should lead to Bragg peaks
unexpected for a strongly pinned Bose-glass. The system under consideration
shares features of the Bragg glass, such as power-law decay of the translation order,
and the strong Bose glass, namely, the diverging tilt modulus  (transverse
Meissner effect). One can expect this behavior for a weak Bose-glass which is
pinned collectively.  Recently, such a glassy phase called Bragg-Bose glass
was observed in numerical simulations of vortices in bulk superconductor at low
concentrations of columnar disorder and low temperatures.\cite{nonomura04,dasgupta05}

A finite temperature   $T>0$ rounds the cusp of the running disorder correlator $R_l(u)$,
so that in the boundary layer $u \sim T_l$, it significantly  deviates form the FP solution
and obeys the  following scaling form:\cite{chauve-00}
\begin{equation} \label{layer}
  R^{\prime\prime}_l(u)=R^{\prime\prime}_l(0)-T_l \left[1-\sqrt{1+(u\chi/T_l)^2} \right],
\end{equation}
where $\chi=|R^{*\prime\prime\prime}(0)|$. However, as was pointed in
Ref.~\onlinecite{chauvembg} the flow equations for columnar disorder have a remarkable
feature in comparison with uncorrelated disorder. Indeed,
substituting the boundary layer scaling (\ref{layer})
in the temperature flow equation~(\ref{ren-T}), we obtain
\begin{equation} \label{T-l}
  \partial_{\ell} T_l = -\theta_T T_l - \varepsilon_d \chi^2/2.
\end{equation}
As follows from Eq.~(\ref{T-l}), the effective temperature $T_l$ vanishes at a
finite length scale $L_{\mathrm{loc}}=e^{l_{\mathrm{loc}}}/\Lambda_0$,
\begin{equation} \label{l-loc}
  l_{\mathrm{loc}} = \frac1{\theta_T}\ln\left(1 + \frac{2\theta_T T_0}{\varepsilon_d
  \chi^2} \right),
\end{equation}
so that the localization effects are settled only on scales
larger than $l_{\mathrm{loc}}>l_c$.

%%%%%%%%%%%%%%%%%%%%%%%%%%%%%%%%%%%%%%%%%%%%%%%%%%%%%%%%%%%%%%%%%%%%%%%%%%%%%%%%%%%%
\begin{table}[tbp]
\caption{Critical exponents for elastic systems with generalized columnar
disorder computed to two-loop order.}
\label{table1}%
\begin{ruledtabular}
\begin{tabular}{llll}
    &                 $\zeta$                                                &  $\psi$           &       $\phi$ \\ \hline
RP  &       $0$                                       &   $\frac{1}{3}\tilde{\varepsilon} + \frac{1}{9}\tilde{\varepsilon}^2$   & $1-\frac{1}{3}\tilde{\varepsilon} + O(\tilde{\varepsilon}^3)$               \\
RF  &   $\frac13 \tilde{\varepsilon}$                 &   $\frac{2}{9}\tilde{\varepsilon} + \frac{5}{162}\tilde{\varepsilon}^2$ & $1-\frac{2}{9}\tilde{\varepsilon} - \frac{1}{18}\tilde{\varepsilon}^2$                           \\
RB  &  {\scriptsize 0.208298}$\tilde{\varepsilon}$    &   {\scriptsize 0.263902}$\tilde{\varepsilon}$                 &  {\scriptsize 1} - {\scriptsize0.263902}$\tilde{\varepsilon}$             \\
    &{\scriptsize\ + 0.006858}$\tilde{\varepsilon}^2 $&   { \ \scriptsize + 0.053615}$\tilde{\varepsilon}^2 $                 & \  - {\scriptsize 0.038941 }$\tilde{\varepsilon}^2 $             \\
\end{tabular}
\end{ruledtabular}
\end{table}
%%%%%%%%%%%%%%%%%%%%%%%%%%%%%%%%%%%%%%%%%%%%%%%%%%%%%%%%%%%%%%%%%%%%%%%%%%%%%%%%%%%%

\subsection{Interacting disordered bosons in (1+1) dimensions}

Let us discuss the special  case of flux lines in (1+1) dimensions which the qualitative
phase diagram is shown in Fig.~\ref{fig-meissner}.   The transverse Meissner physics
for collectively pinned weak Bose glass and small tilt
angles $\vartheta=B_{\pp}/B_{\pr}$ can be explored
using the results obtained in the previous section for the
RP universality class with $\tilde{\varepsilon}=4-2+1=3$. Here, we restore the
dependence on the flux line density $n_0$ fixing the period of $R(u)$ to $1/n_0$.
In contrast to the Bragg-glass,
the weak Bose-glass survives in $d=2$. Indeed, for uncorrelated disorder in $d=2$,
the temperature turns out to be marginally relevant, so that the system has
a line of FPs describing a super-rough phase  with anomalous
growth of the two-point correlation
$\overline{(u_x-u_0)^2}=A(T)\ln^2 x + O(\ln x)$.\cite{schehr07}
According to Eqs.~(\ref{T-l}) and (\ref{l-loc}) for columnar disorder
the temperature vanishes at finite, though a very large scale
\begin{equation}
  l_{\mathrm{loc}}=\frac{2T_0}{n_0^2\chi^2}.
\end{equation}
Unfortunately, the large value of
$\tilde{\varepsilon}$ makes estimation of $\phi$ extremely unreliable. Indeed, the
expansion in $\tilde{\varepsilon}$ shown in Table~\ref{table1} leads to  a zero
value of $\phi$. The exact scaling relation (\ref{phi}) with $\psi$ computed using
the expression from Table~\ref{table1} gives
\begin{eqnarray}
  \phi (1+1) &=& \frac12   \ \ \ \ \mathrm{(one\ loop),} \nonumber \\
  \phi (1+1)&=& \frac13 \ \ \ \ \mathrm{(two\ loop).} \label{phi-frg}
\end{eqnarray}
The one-loop result reproduces the estimation $\phi=1/2$
given by heuristic random walk arguments based on the entropy of flux lines
wandering in the presence of thermal fluctuations.\cite{hwa93} The model
of vortices wandering in a random array of columnar defects can be mapped onto
a quantum  problem of disordered bosons.\cite{nelson92} One can regard each vortex as an
imaginary time world line of a boson, so that the columnar pins parallel to vortices
become quenched pointlike disorder in the quantum problem. The transverse magnetic field
$H_{\pp}$ will play the role of an imaginary vector potential $h$ for the
bosons,\cite{affleck04} so that the bosonic  Hamiltonian turns out to be non-Hermitian:
\begin{eqnarray}
  \hat{\mathcal{H}}&=& - \frac{\hbar^2}{2m} \int dx
  \psi^{\dagger}(x)\left(\partial_x-h\right)^2\psi(x) +
\int dx V(x) \hat{n}(x)
  \nonumber \\
&&  + \frac12 \int dx dx' \hat{n}(x)U(x-x')\hat{n}(x'). \label{H-boson}
\end{eqnarray}
Here, $\psi^{\dagger}(x)$, $\psi(x)$ are the bosonic creation and annihilation operators,
and $\hat{n}(x)=\psi^{\dagger}(x)\psi(x)$ is the density operator.
$U(x)$ is a short-range repulsive interaction potential between bosons
with the strength $U_0=\int dx\,U(x)$. The disorder is described by
a time-independent Gaussian random potential $V(x)$ with zero mean $\overline{V(x)}=0$
and short-range correlations $\overline{V(x)V(x')}=V_0\delta(x-x')$.
We can pass to a quantum hydrodynamic formulation of model (\ref{H-boson})
expressing everything though the bosonic fields $\theta(x)$ and $\varphi(x)$
which satisfy the canonical commutation relation\cite{haldane81}
\begin{equation} \label{comm}
  [\partial_x \varphi(x), \theta(x')] =- i  \delta(x-x').
\end{equation}
For bosons with average density $n_0$, this gives
\begin{eqnarray}
 \hat{ \mathcal{H}} &=& \frac{v_p \hbar}{2} \int dx \left[
 \frac{g}{\pi}  \left(\partial_x \varphi(x)-i h\right)^2 +
 \frac{\pi}g \left(\partial_x \theta(x)\right)^2
 \right]  \nonumber \\
&& + n_0 \int d x\, V(x) \sum_{p=-\infty}^{\infty} e^{2\pi i p (n_0 x + \theta(x))},
\label{H-bos}
\end{eqnarray}
where in the disorder part we have retained only the leading contributions
coming from the backscattering on impurities. The forward scattering
term can be eliminated
by a shift of the phonon field $\theta(x)$ which does not depend on the time $t$,
and thus, this term does not contribute to the current $J \sim \partial_t \theta(x)$.
The Luttinger liquid parameter $g$ and the phonon velocity $v_p$ are given by
\begin{equation}
  g^2  = \frac {\pi^2 n_0 \hbar^2 }{m U_0}, \hspace{11mm}
 v_p^2 = \frac{U_0n_0}{m}.
\end{equation}
The imaginary time ($\tau = i t$) action corresponding to Hamiltonian (\ref{H-bos})
for a particular distribution of disorder can be derived using the
canonical transformation
\begin{eqnarray} \label{S-V}
  \mathcal{S}_V=\int dx d\tau [\mathcal{H}-i\Pi_{\theta}(x)\partial_{\tau}\theta(x)].
\end{eqnarray}
Here, $\partial_{\tau}\theta(x)=(i/\hbar)[H,\theta(x)]= - i gc\Pi_{\theta}(x)/(\pi \hbar)$
and $\Pi_{\theta}=\hbar \partial_x\varphi(x)$ is
the momentum conjugate to $\theta(x)$ which is given by Eq.(\ref{comm}).
Averaging $e^{-S_V/\hbar}$ over disorder by means of the replica trick and
keeping only the most relevant terms, we obtain the replicated action
\begin{eqnarray}
 \mathcal{S} &=&
  \sum\limits_a \int dx d\tau \left\{ \frac{\pi \hbar}{2g} \left[\frac1{v_p}
  \left(\partial_{\tau}\theta_a(x,\tau)\right)^2 \right.\right.  \nonumber \\
  &&+\left. v_p(\partial_{x}\theta_a(x,\tau))^2 \Big]
 + h \partial_{\tau}\theta_a(x,\tau) \right\} - \frac{V_0 n_0^2}{\hbar} \nonumber \\
 && \times \sum\limits_{ab}\int dx d\tau d\tau'
 \cos\left\{2\pi[\theta_a(x,\tau)-\theta_b(x,\tau')]\right\}.\ \ \ \ \ \ \ \label{act-im}
\end{eqnarray}
The imaginary time action (\ref{act-im}) is identical to the Hamiltonian of
periodic elastic system with columnar disorder (\ref{H-rep0}). The imaginary time plays
the role of the longitudinal coordinate $\tau \longleftrightarrow x_{\pr}$ which is parallel
to columnar pins. The Planck constant
stands for the temperature $\hbar \longleftrightarrow T$, and the phonons are related to
the dimensionless displacements field $\theta(x,\tau) =-n_0 u(x)$.
There is the following correspondence between quantities in
the vortices and bosons problems\cite{affleck04}
\begin{eqnarray}
&& g=\frac{\pi T n_0^2}{\sqrt{c_{\pr}c_{\pp}}}, \  \ \ \
 v_p = \sqrt{c_{\pp}/c_{\pr}}.
\end{eqnarray}
The vortex tilt angle $\vartheta$ caused by the transverse field $H_{\pp}$
corresponds to the boson current $J=(-i)\partial \mathcal{H}/\partial h$ induced
by the imaginary vector potential $h$.
For $h=0$, the disordered bosons undergoes a superfluid-insulator transition
at $g=3/2$. This determines the temperature $T_{\mathrm{BG}}$, such that
$g(T_{\mathrm{BG}})=3/2$, above which vortices form a liquid (see Fig.~\ref{fig-meissner}).
It is known that in one dimension there is no difference between bosons and fermions,
and both types of particles are described by the Luttinger liquid  (\ref{H-bos}).
In particular, the hard-core bosons can be mapped onto free fermions that corresponds to
a special value of the Luttinger parameter $g(T^*)=1$, which defines the temperature $T^*$.
In Ref.~\onlinecite{refael06}, the mapping onto free fermions was used to study the
transverse Meissner effect in (1+1) dimensions.
The free fermions on a lattice is described by the tight-binding model,
\begin{eqnarray}
 \hat{\mathcal{H}} = \sum\limits_i \left[ - w_i(c_i^{\dagger} c_{i+1}e^{-h}
  + {c}_{i+1}^{\dagger} {c}_ie^{h})  +(\epsilon_i-\mu) c_i^{\dagger}c_i\right],
   \nonumber  \\
\label{fermions}
\end{eqnarray}
where $c^{\dagger}$, $c$ are on site fermion creation and annihilation operators,
and $\mu$ is the chemical potential. $w_i$ is a random hopping matrix element and
$\epsilon_i$ is a random pinning energy. In Ref.~\onlinecite{refael06}, both cases, the
random pinning and the random hopping models, were studied using the exact
results for the Lloyd model and the strong-randomness real-space RG, respectively.
It was found in both cases that $J\sim h-h_c$, i.e.,  $\phi=1$, that significantly differs
from the FRG prediction (\ref{phi-frg}).
The difference can be attributed to that the free fermions  analog  is limited
to a special point $g=1$ ($T=T^*$),  while the FRG prediction may be valid only for
low temperatures since it is controlled by the zero-temperature fixed point.
The correspondence between the temperature and the Planck constant in both problems
reflects that the zero-temperature FRG FP may have a counterpart in the quantum
problem in the form of an instanton  solution. This may account for the
consistency of the exponent $\phi$ computed by FRG
and estimated using heuristic arguments of kink statistics.

The high-$T_c$ superconductor films grown by deposition often exhibit larger critical
currents than their bulk counterparts due to the formation of dislocations running
parallel to the crystalline axis, and thus, they are natural candidates
to verify the above results. However, as was discussed in Ref.~\onlinecite{rodriguez07},
the picture may be more involved  since the dislocation lines
can meander or they can be of relatively short length
that breaks up the Bose glass  into pieces along the direction of the crystalline axis.

\subsection{Response in the presence of long-range  correlated disorder }
\label{sec42}

We now consider the response to tilt in the presence of  isotropic
LR-correlated disorder. In contrast to the case of generalized columnar disorder
the  $\tilde{h}$ term is not generated due to the analyticity of the LR part $R_2(u)$
of disorder correlator. Moreover, the elasticity remains finite along the FRG flow
though it grows as a power law $c \sim L^{-\psi}$ with $\psi<0$ given by Eq.~(\ref{psi-2})
and shown in Fig.~\ref{fig-psi}. As a consequence, there is no threshold transverse
field: the systems is tilted for any finite tilting force. Renormalizing the balance
equation  $h=c_1\vartheta$ up to the scale $\xi_{\vartheta}$ given by Eq.~(\ref{xi-def}),
we see that the response to the tilting force $h$ is given by a power law
\begin{equation}
  \vartheta(h) \sim h^{\phi},
\end{equation}
with the exponent $\phi>1$ defined by Eq.(\ref{phi}).
The response to  tilt in systems with uncorrelated, columnar and LR-correlated
disorders is shown in Fig.~\ref{fig-resp}.
As one can see from the figure, the response
of systems with LR-correlated disorder interpolates between the
response of systems with uncorrelated and columnar disorder.
In particular, we argue that in the presence of LR-correlated disorder, vortices
can form a new vortex glass phase which exhibits Bragg peaks and vanishing
linear tilt modulus without  transverse Meissner effect.
We will refer to this phase as the strong Bragg glass.

In analogy with the Bose glass, one can attempt to map the system with linear defects
of random orientation corresponding to LR-correlated disorder with
$a=d-1$ to a quantum system consisting of interacting bosons and heavy particles
moving with random quenched velocities according to classical mechanics.

\section{ Kardar-Parisi-Zhang equation with temporally correlated noise  }
\label{sec5}

In this section, we address the relevance of our results to
the Kardar-Parisi-Zhang (KPZ) equation (and closely related Burgers equation),
which describes the dynamics of a
stochastically growing interface.\cite{kardar86}
The latter is characterized by a height function $h(x,t)$, $x\in \mathbb{R}^{d'}$ which
obeys the nonlinear stochastic  equation of motion
\begin{equation} \label{KPZ}
  \partial_t h = \nu \nabla^2 h + \frac{\lambda}2 (\nabla h)^2 + \eta(x,t).
\end{equation}
The first term in Eq.~(\ref{KPZ}) represents the surface tension, while the second
term describes tendency of the surface locally grow to normal itself.
The stochastic noise $\eta(x,t)$ is usually assumed to be Gaussian with
short-range correlations. Here, we consider the noise with long-range correlations in both
time and space. It is defined  in Fourier by\cite{kpzcorr}
\begin{equation}
  \langle \eta(k,\omega)\eta(k',\omega') \rangle =
  2 D(k,\omega) \delta^{d'}(k+k')\delta(\omega+\omega'),
\end{equation}
with the noise spectral density function having power-law singularities of the form
\begin{equation} \label{noise}
  D(k,\omega)= D_0 + D_{\theta} k^{-2\rho}\omega^{-2\theta}.
\end{equation}
Such temporal correlations can originate from impurities which do not diffuse and
impede the growth of the interface, while the space correlations can be due to the presence
of extended defects. Since there is no intrinsic length scale in the problem, asymptotics of
various correlation functions are given by simple power laws.
For instance,  the height-height correlation function scales like
\begin{equation} \label{anzats}
\langle [h(x,t)-h(x',t')]^2 \rangle \sim |x-x'|^{2 \chi} f\left(\frac{|t-t'|}{|x-x'|^z} \right),
\end{equation}
where $\chi$ is the roughness exponent  and $z$ is the dynamic exponent which
describes the scaling of the relaxation time with length
(do not mix it with the dynamic exponent $z$ at the depinning transition,
which is not used in this paper).

Medina \textit{et al}.\cite{kpzcorr} studied the KPZ equation
with the noise spectrum (\ref{noise}) using the dynamical renormalization group (DRG)
approach and here we adopt the notation introduced in their work.
Let us briefly outline the results obtained in Ref.~\onlinecite{kpzcorr}
restricting ourselves mainly to the case $d'=1$.
The flow equations expressed in terms of dimensionless parameters
$U_0=K_{d'}\lambda^2D_0/\nu^3$ and  $U_{\theta}=K_{d'}\lambda^2D_{\theta}/\nu^3$ to
one loop order read
\begin{eqnarray}
&& \!\!\!\!\!\!\!\!\!\!\!\!   \partial_{\ell} \ln \nu =z-2 + \frac{U_0}4  +
  \frac{U_{\theta}}4 (1+2\rho)(1+2\theta)\sec(\pi \theta), \label{ren-nu} \\
&& \!\!\!\!\!\!\!\!\!\!\!\!   \partial_{\ell} \ln \lambda = \chi+ z-2 +
  U_{\theta} \theta (1+2\theta)\sec(\pi \theta), \label{ren-lambda} \ \ \\
&&  \!\!\!\!\!\!\!\!\!\!\!\!   \partial_{\ell} U_{\theta} = U_{\theta}
 [z(1+2\theta)-2\chi-1+2\rho], \label{ren-U-th} \\
&& \!\!\!\!\!\!\!\!\!\!\!\!   \partial_{\ell} U_0 = U_0 (z-2\chi-1)+
  \frac{U_0^2}4 +   \frac{U_{\theta}^2}2 (1+4\theta)\sec(2\pi \theta)\nonumber \\
&& + \frac12 U_0U_{\theta}(1+2\theta)\sec(\pi \theta). \label{ren-U-0}
\end{eqnarray}
Note that the DRG calculations are uncontrolled, in the sense that there
is no small parameter.
For white noise  (${\theta}=0$), the KPZ equation is
invariant under tilting of the surface by a small angle. The STS symmetry
implies that the vertex $\lambda$ does not get corrected by the noise to all orders.
This results  in the exact identity
\begin{equation}
  \chi_{\mathrm{SR}} + z_{\mathrm{SR}} = 2. \label{kpz-rel1}
\end{equation}
Besides the known SR FP with $U_{\theta}=0$, the flow equations  (\ref{ren-nu})-(\ref{ren-U-0})
are expected to have a different LR FP with $U_{\theta}\ne 0$.
It was argued that the term $U_{\theta}$ in the noise spectrum $D(k,\omega)$ acquires
no fluctuation corrections:   the scaling of  $U_{\theta}$ is completely  determined
by its bare dimension so that Eq.~(\ref{ren-U-th}) is exact to all orders.\cite{kpzcorr}
This allows one to compute the exact critical value
$\theta_c=1/6$ (for $\rho=0$) at which there is a crossover from the  SR FP to the LR FP.
For arbitrary $\rho$, the crossover to the LR FP happens at
\begin{equation}
  6\theta+4\rho>1. \label{rel-1}
\end{equation}
The term $U_{\theta}$ becomes relevant and as follows from
Eq.~(\ref{ren-U-th}) the exact relation
\begin{equation} \label{kpz-rel2}
  z_{\mathrm{LR}}(1+2\theta)-2\chi_{\mathrm{LR}}+2\rho=1
\end{equation}
holds at the LR FP.
Let us for the moment ignore the noise correction to $\lambda$ in Eq.~(\ref{ren-lambda}).
This approximation restoring the STS is valid only for small $\theta$ and yields
\begin{eqnarray}
  z^*(\theta,\rho)&=&2-\frac{1+4\theta+2\rho}{3+2\theta}, \label{zz-1}\\
  \chi^*(\theta,\rho)&=&\frac{1+4\theta+2\rho}{3+2\theta}. \label{chi-1}
\end{eqnarray}
For large $\theta$, one can expect a significant deviation of exponents $z$ and $\chi$
from $z^*$ and $\chi^*$. To gain insight into the problem the authors of
Ref.~\onlinecite{kpzcorr} solved
the flow equations (\ref{ren-nu})-(\ref{ren-U-0}) for finite $\theta$ and
$\rho=0$ numerically. They found that the physical LR FP exists only for $\theta<1/4$, while
nothing special is physically expected at $\theta=1/4$.
It was argued that the problem is originated from infrared divergences of
integrals and that infinite number of additional terms generated in the noise spectral density
under DRG:
\begin{equation}
 D(\omega)=\sum_{n=1}^{\infty} D_n\omega^{-2\theta-(n-1 )}.
\end{equation}
Keeping track of renormalization of all $D_n$,
the authors of Ref.~\onlinecite{kpzcorr} solved the truncated system of
flow equations numerically and found that the critical exponents for
$\rho=0$ can be fitted to
\begin{eqnarray}
  \chi(\theta) &=&1.69\theta +0.22, \label{chi-2}\\
  z(\theta)&=&\frac{2\chi(\theta)+1}{1+2\theta}. \label{zz-2}
\end{eqnarray}

We now revise the problem  in the light of what has been learned in the
previous sections.
Using the well-known Cole-Hopf transformation
$Z=\exp[(\lambda/2\nu)h]$ one can  eliminate the nonlinear term in  Eq.~(\ref{KPZ})
and obtain a diffusion equation in time-dependent random potential
\begin{equation} \label{Z-eq}
  \partial_t Z(x,t)= \nu \nabla^2 Z(x,t) + \frac{\lambda}{2\nu}\eta(x,t)Z(x,t).
\end{equation}
The solution of Eq.~(\ref{Z-eq}) can be regarded as the partition function of
a directed polymer (DP) of length $t$ in $(d'+1)$ dimensions with ends fixed
at $(0,0)$ and $(x,t)$:
\begin{eqnarray} \label{DP}
Z(x,t)&=&\int_{x(0)=0}^{x(t)=x} D x(t) \exp\left\{ - \frac1T
\int_0^t dt  \left[ \frac{c}2 (\nabla x(t))^2 \right.\right.\nonumber \\
&& + \eta(x(t),t)\Big] \Big\},
\end{eqnarray}
with $\nu=T/2c$ and $\lambda=1/c$. The DP is
a one-dimensional ($d=1$,$\varepsilon=3$) elastic object with $d'=N$ - dimensional
target space.  Thus, the time-dependent noise $\eta(x,t)$ in the KPZ equation
is mapped to the quenched disorder $V$ in the DP picture.
This gives  the exact relation between the dynamic exponent of KPZ problem and
the DP roughness exponent which reads
\begin{equation}
  z(d')=1/\zeta(d=1,N=d'). \label{z-zeta}
\end{equation}
Spatial correlations in $\eta(x,t)$ corresponds to correlations of quenched disorder
$V$ in the directions transverse to the DP. As the exponent $\rho$ varies
from 0 to 1, the quenched disorder interpolates between RB and RF universality classes.
For example, the exponent $z$ changes from $3/2$ to $1$ for $d'=1$ and white random
noise ($\theta=0$). The stability criterion assures that
the LR FP in the FRG picture is stable if $\zeta_{\textrm{LR}}>\zeta_{\textrm{SR}}$.
This implies that the noise temporal correlations in surface growth problem are relevant
only if the corresponding dynamic exponents fulfill the condition
$z_{\textrm{LR}}<z_{\textrm{SR}}$.
Note that this criterion is purely based on the mapping between the DP and KPZ
problems. Since  $z_{\textrm{SR}}(d'=1,\rho=0)=3/2$ the exponent
(\ref{zz-2}) computed using the modified DRG  violates the criterion of the LR FP
stability,  and thus is ruled out.
Substituting the roughness exponents computed using FRG for the RB  ($\rho=0$) and  RF ($\rho=1$)
universality classes into Eq.~(\ref{z-zeta}) and relating $\delta=3+2\theta$, we obtain the
exact (for $\rho=0,1$ and presumably for any $\rho$) identity
\begin{equation} \label{zz-3}
  z_{\textrm{LR}}=\frac{5-2\rho}{3+2\theta+\psi}.
\end{equation}
%%%%%%%%%%%%%%%%%%%%%%%%     Figure6. Dynamic exponent for KPZ equation %%%%%%%%%%%%%%%%
\begin{figure}[tbp]
\includegraphics[clip,width=3.2 in]{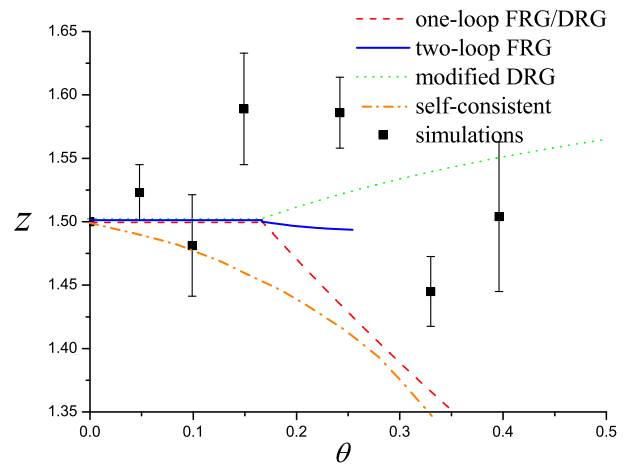}
\caption{(Color online) Dynamic exponent $z$ for the  KPZ equation ($d'=1$)
with temporally correlated noise ($\rho=0$) computed using different technique:
one-loop FRG/DRG is given by Eq.~(\ref{zz-1}); two-loop FRG is given by
Eq.~(\ref{zz-3}) (note that large value ${\varepsilon}=3$ does not
allow for accurate computation of $\psi$ and $z$ at large $\theta$);
modified DRG is given by Eq.~(\ref{zz-2}); self-consistent  approximation
of Ref.~\onlinecite{katzav04} and simulations of Ref.~\onlinecite{lam92}.}
\label{fig-kpz}
\end{figure}
%%%%%%%%%%%%%%%%%%%%%%%%%%%%%%%%%%%%%%%%%%%%%%%%%%%%%%%%%%%%%%%%%%%%%%%%%%%%%%%%
To one-loop order in FRG, i.e., for $\psi=0$, exponent (\ref{zz-3}) coincides with
the estimation given by DRG (\ref{zz-1}) for small $\theta$.
Though the exponent $\psi$ has been computed in Sec.~\ref{sec33}
for the RB  ($\rho=0$) and  RF ($\rho=1$) universality classes to two-loop order
in a controllable way, the large value $\varepsilon=3$ of the expansion
parameter describing the DP problem makes the estimation of $\psi$ highly unreliable.
Nevertheless, since $\psi = 0$ is zero on
the crossover line between the LR and SR FPs, we can determine this line exactly for
$\rho=0,1$ from equation $z_{\textrm{LR}}<z_{\textrm{SR}}=3/2$
that leads back to Eq.~(\ref{rel-1}).
Taking into account that $\psi$ is nonpositive for columnar disorder, we obtain
the lower and upper bounds on $z(\theta)$ for $\rho=0$ and
$\theta\in[\frac16,\frac12]$ as
\begin{equation}
 \frac{5}{3+2\theta} \le z(\theta) \le \frac32. \label{bounds}
\end{equation}
The critical exponent $z$ computed using FRG, DRG, and measured in numerical
simulations of Ref.~\onlinecite{lam92} is shown in Fig.~\ref{fig-kpz}.
The KPZ equation with temporally correlated noise was also studied using a self-consistent
approximation (SCA).\cite{katzav04}
The SCA equations have two strong-coupling solutions. The first one
exhibits  a crossoverlike behavior at $\theta=\frac16$ and corresponds to the
one-loop FRG prediction.
The second solution, which is considered to be dominant, leads to
a smooth dependence of $z$ on $\theta$ shown in Fig.~\ref{fig-kpz}.
Both the  SCA solutions are in agreement with the FRG prediction that the
exponent $z$ is a decreasing function of $\theta$, while the modified DRG suggests that
$z$ increases with $\theta$. However, the second SCA solution considered to be dominant
does not satisfy bounds (\ref{bounds}), and thus is ruled out.

Let us generalize identity (\ref{kpz-rel1}) to the case of temporally correlated noise.
Note that the solution of the KPZ equation $h(x,t)$ gives the free energy of DP (\ref{DP}).
The free energy per unit length $f(\vartheta)$ of the DP tilted by the
transverse field $H_{\pp}$ to the angle $\vartheta$ can be written as
$f(\vartheta)=f(0)+ \tilde{c} \vartheta^{\alpha} -H_{\pp}\,\vartheta$.
The naive elastic approximation suggests $\alpha=2$. In order to take into account the
renormalization of  elasticity, we determine the exponent $\alpha$ from the condition
that at equilibrium the response to the field $H_{\pp}$ is
$\vartheta\sim H_{\pp}^{\phi}$. This
fixes $\alpha=1+1/\phi$ with $\phi$ given by Eq.~(\ref{phi}).
Then the total free energy of the DP of length $t$
can be written as a function of the free end coordinate $x$  as follows:
\begin{equation} \label{last}
  h(x)= t f(0) +tc(x/t)^{\alpha} - x^{\chi}.
\end{equation}
The last term in Eq.~(\ref{last}) describes the typical fluctuation of the free energy
due to the disorder and is given by Eq.~(\ref{anzats}). Balancing the last two
terms of Eq.~(\ref{last}) and using Eq.~(\ref{phi}), we obtain the exact scaling relation
\begin{equation}
  \chi+z=2-\psi z, \label{new-rel}
\end{equation}
which holds at the LR FP as well as at the SR FP. At the SR FP $\psi=0$, so that
Eq.~(\ref{new-rel}) reduces to the STS identity (\ref{kpz-rel1}). Excluding
$\psi$ from Eqs.~(\ref{new-rel}) and (\ref{zz-3}), we arrive at the
relation~(\ref{kpz-rel2}) valid at the LR FP.

\section{Summary}
\label{sec6}

We have studied the large-scale behavior of elastic systems such as interfaces and lattices
pinned by correlated disorder using the functional renormalization group.
We consider two types of disorder correlations: columnar disorder
generalized to  extended defects
and LR-correlated disorder. Both types of disorder correlations can be produced
in real systems, for example, by subjecting them to either static or rotating
ion beam irradiation.
We have computed the critical exponents to second order
in $\varepsilon=4-d$ and $\delta=4-a$ for LR-correlated disorder and to second
order in $\tilde{\varepsilon}=4-\varepsilon+\varepsilon_d$ for $\varepsilon_d$-dimensional
extended defects. The correlation of disorder violates the statistical tilt symmetry
and results in a highly nonlinear response to a tilt.
In the presence of generalized columnar disorder, elastic systems exhibit a transverse
Meissner effect: disorder generates the critical field $h_c$ below which
there is no response to a tilt and above which the tilt angle behaves
as $\vartheta\sim(h-h_c)^{\phi}$ with a universal exponent $\phi<1$.
The periodic case describes a weak Bose glass which is expected in type-II superconductors
with columnar disorder at small temperatures and at high vortex density which exceeds
the density of columnar pins. The weak Bose glass is pinned collectively and
shares features of the Bragg glass, such as a power-law decay of translational order,
and features of the strong Bose glass, such as a transverse Meissner effect.
For isotropic LR-correlated disorder, the linear tilt modulus vanishes
at small fields leading to a power-law response $\vartheta\sim h^{\phi}$
with $\phi>1$.
The response of systems with LR-correlated disorder interpolates between
the response of systems with uncorrelated and columnar disorder.
We argued that in the presence of LR-correlated disorder vortices
can form a strong Bragg glass which exhibits Bragg peaks and a vanishing
linear tilt modulus without transverse Meissner effect.
The elastic one-dimensional interface, i.e., the directed polymer,
in the presence of LR-correlated disorder
can be  mapped to the Kardar-Parisi-Zhang equation with temporally correlated noise.
Using this mapping, we have computed the critical exponents describing the surface growth and
compared with the exponents obtained using dynamical renormalization group,
self-consistent approximation, and numerical simulations.

\begin{acknowledgments}
I would like to thank Kay Wiese, Pierre Le Doussal, Joachim Krug, and David Nelson
for inspiring discussions, and the Max Planck Institute for
the Physics of Complex Systems in Dresden for hospitality, where part of
this work was done.
This work has been supported by the European Commission under contract No.
MIF1-CT-2005-021897 and partially by the
Agence Nationale de la Recherche (05-BLAN-0099-01).
\end{acknowledgments}

\appendix

\section{Correction to elasticity: two-loop diagrams}
\label{app-A}

In this  Appendix,  we calculate diagrams shown in Fig.~\ref{fig-2loop}
keeping only the terms which correct the elasticity.
Here, we  set $g_1(x):=\delta^d(x)$ and $g_2(x):=g(x)$.
Diagram $a$ yields
\begin{eqnarray}
  [a]_{\alpha \beta} &=& - \frac1{2T} \int_{y_1,y_2,y_3}
  R_{\alpha}^{\prime\prime}(u_x-u_{x-y_1-y_2-y_3})  \nonumber \\
 &&\times R_{\beta}^{(4)}(u_{x-y_1}-u_{x-y_1-y_2})
  g_{\alpha}(y_1+y_2+y_3)  \nonumber \\
 &&\times  g_{\beta}(y_2) \prod\limits_{l=1}^3 C(y_l),
\end{eqnarray}
where $\alpha,\beta=1,2$ and $C(x)$ is given by Eq.~(\ref{C}).
Using the short distance expansion (\ref{short}), we obtain
\begin{eqnarray}
  [a]_{\alpha \beta}^{(1)} &=& -\frac1{4dT} (\nabla u_x)^2
   R^{(4)}_{\alpha}(0) R^{(4)}_{\beta}(0)
  \int_{\{y\}} \prod\limits_{l=1}^3 C(y_l) \nonumber \\
  &&\times
  (y_1+y_2+y_3)^2 g_{\alpha}(y_1+y_2+y_3) g_{\beta}(y_2)\ \ \
\end{eqnarray}
and
\begin{eqnarray}
  [a]_{\alpha \beta}^{(2)} &=& -\frac1{4dT} (\nabla u_x)^2
   R^{\prime\prime}_{\alpha}(0) R^{(6)}_{\beta}(0)
  \int_{\{y\}} \prod\limits_{l=1}^3 C(y_l) \nonumber \\
  &&\times
  y_2^2 g_{\alpha}(y_1+y_2+y_3) g_{\beta}(y_2).\ \ \
\end{eqnarray}
Note that the term $R^{\prime\prime\prime}_{\alpha}(0) R^{\prime\prime\prime}_{\beta}(0)$
does not contribute since LR disorder $R_2(u)$ remains an
analytic function along the FRG flow, while all diagrams with $\alpha=\beta=1$ are zero due
to the STS. We will neglect similar terms in what follows.
For diagram $b$, we have
\begin{eqnarray}
  [b]_{\alpha \beta} &=&  \frac1{2T} \int_{y_1,y_2,y_3}
  R_{\alpha}^{\prime\prime}(u_{x-y_1}-u_{x-y_2})  \nonumber \\
 &&\times R_{\beta}^{(4)}(u_{x}-u_{x-y_3})
  g_{\alpha}(y_2-y_1)  \nonumber \\
 &&\times  g_{\beta}(y_3) \prod\limits_{l=1}^3 C(y_l).
\end{eqnarray}
Applying the short distance expansion (\ref{short}), we arrive at
\begin{eqnarray}
  [b]_{\alpha \beta}^{(1)} &=& \frac1{4dT} (\nabla u_x)^2
   R^{(4)}_{\alpha}(0) R^{(4)}_{\beta}(0)
  \int_{\{y\}} \prod\limits_{l=1}^3 C(y_l) \nonumber \\
  &&\times
  (y_1+y_2)^2 g_{\alpha}(y_1+y_2) g_{\beta}(y_3)\ \ \
\end{eqnarray}
and
\begin{eqnarray}
  [b]_{\alpha \beta}^{(2)} &=& \frac1{4dT} (\nabla u_x)^2
   R^{\prime\prime}_{\alpha}(0) R^{(6)}_{\beta}(0)
  \int_{\{y\}} \prod\limits_{l=1}^3 C(y_l) \nonumber \\
  &&\times
  y_3^2 g_{\alpha}(y_1+y_2) g_{\beta}(y_3).\ \ \
\end{eqnarray}
Diagram $[c]$ gives
\begin{eqnarray}
  [c]_{\alpha \beta} &=& - \frac1{2T} \int_{y_1,y_2,y_3}
  R_{\alpha}^{\prime\prime\prime}(u_{x}-u_{x-y_1-y_2})  \nonumber \\
 &&\times R_{\beta}^{\prime\prime\prime}(u_{x-y_1}-u_{x-y_1-y_2-y_3})
  g_{\alpha}(y_1+y_2)  \nonumber \\
 &&\times  g_{\beta}(y_2+y_3) \prod\limits_{l=1}^3 C(y_l).
\end{eqnarray}
After short distance expansion, we find that diagrams $[c]$ give rise to elasticity correction
only for  $\alpha=\beta=2$ which reads
\begin{eqnarray}
  [c]_{\alpha \beta} &=& - \frac1{2dT} (\nabla u_x)^2
   R^{(4)}_{\alpha}(0) R^{(4)}_{\beta}(0)
  \int_{\{y\}} \prod\limits_{l=1}^3 C(y_l) \nonumber \\
  &&\!\!\!\!\!\!\!\!\!\! \!\!\!\!\!\! \times
  [(y_1+y_2)\cdot(y_2+y_3)] g_{\alpha}(y_1+y_2) g_{\beta}(y_2+y_3).\ \ \
\end{eqnarray}
Straightforward analysis shows that
\begin{eqnarray}
&&    [a]_{1 \alpha}^{(1)} = [a]_{\alpha 1}^{(2)} = [b]_{1 \alpha}^{(1)}
  =[b]_{\alpha 1}^{(2)} = 0, \ \ \ \ (\alpha=1,2), \nonumber \\
&& [a]^{(1)}_{21}+[b]^{(1)}_{21}=0, \nonumber \\
&& [c]_{11}=[c]_{12}=[c]_{21}=0.
\end{eqnarray}
We now compute the integrals combining them in pairs:
\begin{eqnarray}
&& \!\!\!\!\!\!\!\!\!\!\!\! [a]^{(1)}_{22}+[b]^{(1)}_{22}
  =  \frac{(a-d)(a-2)c}{4dT \hat{m}^{2\delta}}
   R_2^{(4)}(0)^2 (\nabla u_x)^2 \nonumber \\
&& \!\!\!\!\!\! \times \int_{\{q\}}
\left[\frac{1}{(q_1+q_2)^2+1} - \frac{1}{q_2^2+1} \right]
\frac{q_1^{a-d-2}q_2^{a-d}}{(q_1^2+1)^2}, \label{A1} \qquad
\end{eqnarray}
where we have included $1/c^2$ in redefinition of $R_i(u)$.
The integral over $q_i$ in Eq.~(\ref{A1}) is of order $O(1/\varepsilon)$ so
that $[a]^{(1)}_{22}+[b]^{(1)}_{22}$ is finite and does not correct
elasticity at two-loop order. Other diagrams give
\begin{eqnarray}
&& [a]^{(2)}_{12}+[b]^{(2)}_{12} = \frac{c(\nabla u_x)^2}{2T}
\frac{(a-d)(a-2)}{2d} \hat{m}^{-(\varepsilon+\delta)} \nonumber \\
&& \hspace{20mm} \times (J_1-J_2)  R_1^{\prime\prime}(0) R_2^{(6)}(0),  \\[3mm]
&& [a]^{(2)}_{22}+[b]^{(2)}_{22} = \frac{c(\nabla u_x)^2}{2T}
\frac{(a-d)(a-2)}{2d} \hat{m}^{-2\delta} \nonumber \\
&& \hspace{20mm} \times (J_3-J_4)  R_2^{\prime\prime}(0) R_2^{(6)}(0),
\end{eqnarray}
where we have defined the following two-loop integrals:
\begin{eqnarray}
&&  \!\!\!\!\!\!\!\! J_1= \int_{q} \frac{q_2^{a-d-2}}{(1+q_1^2)^2[1+(q_1+q_2)^2]} =
\frac{K_4^2}{\delta(\delta+\varepsilon)} + O\left({\varepsilon}^{-1},\delta^{-1}\right),
 \nonumber \\
&& \!\!\!\!\!\!\!\! J_2= \int_{q} \frac{q_2^{a-d-2}}{(1+q_1^2)^2(1+q_2^2)} =
 \frac{K_4^2}{\delta\, \varepsilon} + O\left({\varepsilon}^{-1},\delta^{-1}\right), \nonumber \\
&& \!\!\!\!\!\!\!\! J_3 = \int_{q} \frac{q_1^{a-d} q_2^{a-d-2}}{(1+q_1^2)^2[1+(q_1+q_2)^2]}
 = \frac{K_4^2}{2 \delta^2} + O\left({\varepsilon}^{-1},\delta^{-1}\right), \nonumber \\
&& \!\!\!\!\!\!\!\!  J_4= \int_{q} \frac{q_1^{a-d} q_2^{a-d-2}}{(1+q_1^2)^2(1+q_2^2)} =
\frac{K_4^2}{\delta^2}+ O\left({\varepsilon}^{-1},\delta^{-1}\right).
\end{eqnarray}
The last diagram
\begin{eqnarray}
&& [c]_{22} =  -\frac{(a-d)^2 c}{2dT \hat{m}^{2\delta}}
   R_2^{(4)}(0)^2 (\nabla u_x)^2 \nonumber \\
&& \times \int_{\{q\}}
\frac{\left[q_1\cdot q_2 \right]
q_1^{a-d-2}q_2^{a-d-2}}{[(q_1+q_2)^2+1](q_1^2+1)(q_2^2+1)} \label{A3} \ \ \
\end{eqnarray}
is finite in the limit  $\varepsilon,\delta \to 0 $, and thus
does not correct elasticity at two-loop order.

\end{document}